\documentclass[twocolumn,epjc3]{svjour3} 
\smartqed 
\usepackage{widetext}
\usepackage[T1]{fontenc}
\usepackage{epsfig,amsmath}
\usepackage{graphicx,undertilde}
\usepackage[colorlinks,citecolor=blue,urlcolor=blue,linkcolor=blue]{hyperref}
\journalname{Eur. Phys. J. C}

\newcommand{\nc}{{\sc ASperGe}}
\newcommand{\gsl}{{\sc Gsl}}

\newcommand{\Tr}{\text{Tr}}
\newcommand{\etc}{{\it etc.}}
\newcommand{\ie}{{\it i.e.}}
\newcommand{\eg}{{\it e.g.}}

\def\be{\begin{equation}}
\def\ee{\end{equation}}

\def\bsp#1\esp{\begin{split}#1\end{split}}
\def\bpm{\begin{pmatrix}} 
\def\epm{\end{pmatrix}} 

\newcommand{\cpp}{{\sc C++}}
\newcommand{\feynrules}{{\sc Feyn\-Rules}}
\newcommand{\feynarts}{{\sc Feyn\-Arts}}
\newcommand{\mathematica}{{\sc Mathematica}}
\newcommand{\calchep}{{\sc Calc\-Hep}}
\newcommand{\comphep}{{\sc Comp\-Hep}}
\newcommand{\sherpa}{{\sc Sher\-pa}}
\newcommand{\whizard}{{\sc Whizard}}
\newcommand{\formcalc}{{\sc Form\-Calc}}
\newcommand{\gosam}{{\sc GoSam}}
\newcommand{\hwpp}{{\sc Herwig++}}
\newcommand{\python}{{\sc Python}}
\newcommand{\madgraph}{{\sc MadGraph}}
\newcommand{\madevent}{{\sc Mad\-E\-vent}}
\newcommand{\madanalysis}{{\sc MadAnalysis}}
\newcommand{\aloha}{{\sc Aloha}}

\begin{document}

\title{Automated mass spectrum generation for new physics}
\author{
   Adam Alloul \thanksref{a1} \and
   Jorgen D'Hondt \thanksref{a2} \and
   Karen De Causmaecker \thanksref{a2} \and 
   Benjamin Fuks \thanksref{a1,a3} \and
   Michel Rausch de Traubenberg \thanksref{a1}
}

 
\institute{Institut Pluridisciplinaire Hubert Curien/D\'epartement Recherches
    Subatomiques, Universit\'e de Strasbourg/CNRS-IN2P3,
    23 Rue du Loess,  F-67037  Strasbourg, France\label{a1}
  \and
  Interuniversity Institute for High Energies (IIHE), ELEM, and Theoretische Natuurkunde, 
  Vrije Universiteit Brussel and The International 
  Solvay Institutes, Pleinlaan 2, B-1050 Brussels, Belgium
  \label{a2}
  \and
  Theory Division, Physics Department, CERN, CH-1211 Geneva 23,
    Switzerland\label{a3}
}

\date{Received: date / Accepted: date}

\maketitle

\vspace*{-80mm}
\noindent IPHC-PHENO-13-01; CERN-PH-TH/2013-012 
\vspace*{73mm}
 
\begin{abstract}
  We describe an extension of the \feynrules\ package dedicated to the automatic generation
  of the mass spectrum associated with any Lagrangian-based quantum field theory. After
  introducing a simplified way to implement particle 
  mixings, we present a new class of \feynrules\ functions allowing both for the
  analytical computation of all the model mass matrices and for the generation
  of a \cpp\ package, dubbed \nc. This program can then be further employed for a numerical
  evaluation of the rotation matrices necessary to diagonalize the field basis.
  We illustrate these features in the context of the Two-Higgs-Doublet Model, 
  the Minimal Left-Right Symmetric Standard Model and the Minimal Supersymmetric Standard Model. 
\end{abstract}


\section{Introduction}
\label{sec:intro}
Although the Standard Model of particle physics is very well verified empirically at 
the current accessible energies, numerous extensions to its Lagrangian are proposed. 
These extensions describe new or alternative fundamental interactions that typically 
accommodate possible new physics phenomena at higher energies as well as at the current 
collider energies. In this top-down approach, the phenomenology of the proposed extensions 
is to be confronted with experimental observations. In order to obtain the mass spectrum 
of any new physics model reflected by its Lagrangian, the mixing matrices of the gauge 
eigenstates into the mass eigenstates are needed. An automated mass spectrum generator, 
\nc\footnote{The acronym \nc\ stands for \textbf{A}utomated \textbf{Spe}ct\textbf{r}um 
\textbf{Ge}neration.} is developed within the framework of the \feynrules\ program 
\cite{Christensen:2008py,%
Christensen:2009jx,Christensen:2010wz,Duhr:2011se,Fuks:2012im} to determine the 
mixing matrices numerically. This allows a study of the direct relation between the
parameters of any new physics model and the observable masses of the fundamental particles. 

This paper describes in Section \ref{sec:feynrules} and \ref{sec:technicalFR} the 
relevant parts of the \feynrules\ program to introduce the new \nc\ package, detailed 
in Section \ref{sec:ass}. To illustrate its application several examples are presented 
in Section~\ref{sec:example}.

\section{The \feynrules\ package}\label{sec:feynrules}
The program \feynrules\ \cite{Christensen:2008py,Christensen:2009jx,%
Christensen:2010wz,Duhr:2011se,Fuks:2012im} is a \mathematica
\footnote{\mathematica\ is a registered trademark of Wolfram Research, Inc.}
package that allows for the automated extraction of Feynman rules from any
Lagrangian describing the dynamics of a perturbative quantum
field theory. The Feynman rules, together with general information such as
the definitions of the model particles or of the Lagrangian
parameters, can subsequently be exported by means of several translation
interfaces to
matrix-element generators. Up to now, interfaces to \comphep\ and \calchep\
\cite{Pukhov:1999gg,Boos:2004kh,Pukhov:2004ca,Belyaev:2012qa},
\feynarts\ and \formcalc\ \cite{Hahn:1998yk,Hahn:2000kx,Hahn:2009bf,%
Agrawal:2011tm},
\madgraph\ and \madevent\ \cite{Stelzer:1994ta,Maltoni:2002qb,Alwall:2007st,%
Alwall:2008pm,Alwall:2011uj}, \sherpa\ \cite{Gleisberg:2003xi,Gleisberg:2008ta}
and \whizard\ \cite{Moretti:2001zz,Kilian:2007gr} have been developed. 

In addition, any model can also be converted to a \python\ library containing 
classes and objects representing particles, parameters and vertices. This
format is dubbed the Universal \feynrules\ Output (UFO) format
\cite{Degrande:2011ua} and is appropriate to address the implementation of any
high-energy physics model into computational tools. Its strength lies in its
agnosticism with respect to the allowed Lorentz
and/or color structures appearing in the Lagrangian, in contrast to any other
more conventional model format for which restrictions are imposed. Presently,
the UFO is used by 
\aloha\ \cite{deAquino:2011ub}, 
\madanalysis\ 5 \cite{Conte:2012fm}, and \madgraph\ 5 \cite{Alwall:2011uj}, 
and will be used, in the future, by \gosam\ \cite{Cullen:2011ac,Cullen:2011xs} 
and \hwpp\ \cite{Bahr:2008pv}.

The \feynrules\ model structure extends the format employed in
\feynarts\ \cite{Hahn:2000kx} so that particles, parameters
and gauge groups are now all defined in a similar fashion. Following the
\feynarts\ conventions, particles are collected
into classes describing multiplets having exactly the same quantum numbers, but
possibly different mas\-ses. Each of these classes is defined as a set of
\mathematica\ replacement rules referring to its properties. For example, the
three vector fields $W_i$ associated to the $SU(2)_L$ gauge subgroup of the
Standard Model could be declared as
\begin{verbatim}
  V[1] == { 
    ClassName     -> Wi,
    Unphysical    -> True,
    SelfConjugate -> True, 
    Indices       -> { Index[SU2W] },
    FlavorIndex   -> SU2W,
    Definitions   -> { 
     Wi[mu_,1]->(Wbar[mu]+W[mu])/Sqrt[2],
     Wi[mu_,2]->(Wbar[mu]-W[mu])/(I*Sqrt[2]), 
     Wi[mu_,3]->cw Z[mu] + sw A[mu]}
  }
\end{verbatim}
This set of \mathematica\ rules defines a vector field
(\texttt{V[1]}) represented by the symbol \texttt{Wi} (its
\texttt{Class\-Name}) and carrying a flavor
index {\tt SU2W} associated with the adjoint gauge
index of $SU(2)_L$. As this field is declared as unphysical\footnote{In this
work, we denote as unphysical any field that is not a mass eigenstate of the
theory.}
(\texttt{Unphysical->True}), it must be linked to one or several of the
mass eigenstates of the model by means of appropriate mixing relations. As
illustrated in the example, these
relations are passed through the attribute \texttt{Definitions} of the particle
class. They can either be purely numerical, as for the $W_1$ and $W_2$ bosons
that are rotated to the charged $W^+$ and $W^-$ bosons, or depend on some of the
model parameters, as for the $W_3$ field which is re-expressed in terms of the
photon $A$ and the $Z$ boson through a relation depending on the
sine and cosine of the electroweak mixing angle (\texttt{sw} and \texttt{cw}). 

In contrast, declaring physical particles requires the implementation of extra
information such as their masses (\texttt{Mass}),
widths (\texttt{Width}) and Particle Data Group codes (\texttt{PDG})
\cite{Beringer:1900zz}. The $Z$-boson field introduced above could hence be
declared as 
\begin{verbatim}
  V[2] == { 
    ClassName       -> Z, 
    SelfConjugate   -> True,
    Mass            -> {MZ, 91.1876},
    Width           -> {WZ, 2.4952},
    PDG             -> 23
  }
\end{verbatim}

The declaration of the model parameters
and gauge group is similar and based on dedicated classes with their own
set of attributes. Since only the particle class properties 
introduced above are sufficient for the understanding of the present work, we
omit any further detail and refer the reader to Refs.\
\cite{Christensen:2008py,Duhr:2011se} for more information on particle,
parameter and gauge group implementation in \feynrules.
 
The last key ingredient to achieve a model implementation consists of its
Lagrangian. It is provided using standard \mathematica\ commands, augmented by
some special symbols representing objects such as Dirac matrices, 
vector field strength tensors or covariant de\-ri\-va\-ti\-ves. 
The user has then the possibility to perform basic checks on the Lagrangian, such
as verifying its hermiticity, the normalization of kinetic terms, \etc\
We again refer to the \feynrules\ manual for more information
\cite{Christensen:2008py}.

After having imported the \feynrules\ package into the current \mathematica\
session, the model is loaded by issuing 
\begin{verbatim}
  LoadModel[ "file1.fr", "file2.fr", ... ]
\end{verbatim}
where its implementation can be possibly spread among the files
\texttt{"file1.fr"}, \texttt{"file2.fr"}, \etc, according to the convenience of the
user. It can also be directly implemented 
within the \mathematica\ session so that the function \texttt{LoadModel} is
called, in that case, without any argument. 

The Feynman rules can be subsequently extracted by means of the command 
\begin{verbatim}
  FeynmanRules[ Lag ]
\end{verbatim}
where \texttt{Lag} is the \mathematica\ symbol containing the expression of the
Lagrangian, written in four-di\-men\-si\-o\-nal spacetime and employing 
four-component spinors for fermions. The \texttt{FeynmanRules}
method extracts the interaction vertices included in the
Lagrangian \texttt{Lag} so that they can be further employed within 
\mathematica\ for dedicated studies. 

All the interfaces to Monte Carlo event generation tools can be invoked in a similar 
procedure, 
\begin{verbatim}
  WriteXXX[ Lag ]
\end{verbatim}
where the sequence of letters \texttt{XXX} takes one of the values
\texttt{CHOutput} (\calchep), \texttt{Feyn\-Arts\-Out\-put} (\feynarts),
\texttt{SHOutput} (\sherpa), \texttt{UFO} or \texttt{WOOutput} (\whizard).

In the context of supersymmetric theories, the most natural
and convenient way to construct a Lagrangian consists of employing the
superspace formalism. Therefore, the \feynrules\ package includes a module
allowing for superfield declarations and La\-gran\-gian implementation in terms
of superfields. Dedicated functions are then provided to convert 
superfield expressions into a form more suitable with respect to the
requirements of the interfaces to the Monte Carlo event generators~\cite{Duhr:2011se}.


\section{Implementing mixings in \feynrules}\label{sec:technicalFR}

In this work, we propose an extension of the \feynrules\ package aiming to
simplify the declaration of the mixing relations linking the unphysical degrees
of freedom of the theory to the physical fields. This new module allows to
automatically fill the 
\texttt{Definitions} attribute of the fields, where relevant, and declare
the mixing matrices as parameters. In addition, we have developed an
interface generating a \cpp\ code dedicated to the
diagonalization of the mass matrices of the model (see Section \ref{sec:ass})
after having implemented in
\feynrules\ a function allowing for their analytical extraction
from the Lagrangian. In this way, the values of all the mixing parameters are
derived numerically and can be re-imported into \feynrules. 

\subsection{Mixing declarations}\label{sec:mixdecl}
For an efficient declaration of the mixing relations, we have extended the
\feynrules\ model file structure by adjoining a new class dedicated to particle
mixings. Consequently, all mixing relations among the states can be declared on
the same spirit as particles, gauge groups and parameters, after having been
gathered into a list dub\-bed \texttt{M\$MixingsDescription}
\begin{verbatim}
  M$MixingsDescription = { 
     Mix["l1"] == { options1 }, 
     Mix["l2"] == { options2 }, 
            ...
  } 
\end{verbatim}
Each element of this list consists of an equality dedicated to one specific mixing
relation. It associates a label, given as a string, (\texttt{"l1"},
\texttt{"l2"}, \etc) with a set of \mathematica\ replacement
rules defining the mixing properties (\texttt{options1}, \texttt{options2},
\etc).  

In order to illustrate the choice of options offered to
the user, we consider the example of Section \ref{sec:feynrules} where we have
focused on the mixing of the $SU(2)_L$ gauge bosons. We start by implementing
the mixing of the $W_1$ and $W_2$ gauge fields,
\be
  W_\mu^+ = \frac{W_\mu^1 - i W^2_\mu}{\sqrt{2}}   \quad\text{and}\quad
  W_\mu^- = \frac{W_\mu^1 + i W^2_\mu}{\sqrt{2}} \ ,
\label{eq:wmix}\ee
which stems from the diagonalization of the third generator of
$SU(2)_L$ in the adjoint representation. As Eq.\ \eqref{eq:wmix} is purely
numerical, \ie, it does not involve any model parameter, it can be declared in
the model file in a very compact form, 
\begin{verbatim}
  Mix["Wmix"] == {
    MassBasis  -> {W, Wbar}, 
    GaugeBasis -> {Wi[1], Wi[2]},
    Value      -> { {1/Sqrt[2], -I/Sqrt[2]}, 
                    {1/Sqrt[2], I/Sqrt[2]}  }
  } 
\end{verbatim}    
The command above declares a mixing relation, dubbed \texttt{Wmix}, that can be 
schematically written as 
\begin{verbatim}
  MassBasis = Value . GaugeBasis
\end{verbatim}
where the dot product stands for the usual matrix product. The information on
the gauge basis is provided as the value of the attribute \texttt{GaugeBasis}
which refers here to the unphysical fields $W_1$ (\texttt{Wi[1]}) and $W_2$
(\texttt{Wi[2]}). Similarly, the \texttt{MassBasis} attribute refers to the mass
basis, containing here the symbols associated with the
$W^+$ (\texttt{W}) and $W^-$ (\texttt{Wbar}) bosons. Finally, the mixing matrix
is given under a numerical form as the argument of the attribute
\texttt{Value}.

Some remarks are in order. First, the gauge basis only contains unphysical
fields, while the mass basis can contain either physical fields, unphysical
fields or both. Particle mixings can therefore be possibly implemented in
several steps, as illustrated in Section \ref{sec:LRSM}. Next, 
spin and Lorentz indices can be omitted and  
the index ordering is defined when declaring 
the fields (through the attribute {\tt Indices} of the particle class
\cite{Christensen:2008py}). Finally, if some indices
are irrelevant, \ie, if they are identical
for all the involved fields, underscores can be employed to simplify the
mixing declaration. 
For instance, the three left-handed
down-type squarks $\tilde d_L^c$ (\texttt{sdL[1,c]}), $\tilde s_L^c$
(\texttt{sdL[2,c]}), and $\tilde b_L^c$ (\texttt{sdL[3,c]}) are related to
the squark gauge-eigenstates $\tilde Q_L^{ifc}$ (\texttt{QLs[i,f,c]}), the index
$i$ being a fundamental $SU(2)_L$ index, the index $f$ a flavor index and the
index $c$ a fundamental color index. The corresponding declaration reads
\begin{verbatim}
  Mix["sdleft"] == { 
    MassBasis  -> 
      {sdL[1,_], sdL[2,_], sdL[3,_]}, 
    GaugeBasis -> 
      {QLs[2,1,_], QLs[2,2,_], QLs[2,3,_]}, 
             ... 
  }
\end{verbatim}
where the mixing matrix is the identity. The underscore reflects that the
same color index is carried by all fields. 

We now get back to weak gauge boson mixings and turn to the neutral sector. We
hence focus on the rotation of the third weak boson $W_3$ and the hypercharge
gauge boson $B$ to the photon and $Z$-boson states, 
\be
  \bpm  A_\mu \\ Z_\mu \epm = U_w \bpm B_\mu \\ W_\mu^3 \epm \ ,
\label{eq:Zmix}\ee
after introducing the \textit{a priori} unknown weak mixing
matrix $U_w$\footnote{Following more standard conventions, the relation of Eq.\
\eqref{eq:Zmix} is usually written in 
terms of the cosine and sine of the electroweak mixing angle, as in Section
\ref{sec:feynrules}. However, we have adopted the choice of staying fully
general for the sake of the example.}. The computation of the numerical values
of its matrix elements is addressed by means of the \cpp\ package generated by
\feynrules\ (see Section \ref{sec:ass}) and is only possible if the mixing is
declared according to the syntax 
\begin{verbatim}
  Mix["AZmix"] == {
    MassBasis    -> {A, Z}, 
    GaugeBasis   -> {B, Wi[3]}, 
    MixingMatrix -> UW, 
    BlockName    -> WEAKMIX
  } 
\end{verbatim}
The declaration of the gauge and mass bases is similar to the case of the
charged $W$ bosons, while the attribute \texttt{Value} has
been removed as the numerical value of the mixing matrix is not known. The user
provides instead the symbol referring to the mixing matrix (\texttt{UW}) by
means of the \texttt{MixingMatrix} attribute, \textit{without} declaring it as
one of the model parameters. This last task is internally handled by \feynrules\
which assumes that the mixing matrix is complex and which creates two external
tensorial parameters, one for the real part and one for the
imaginary part of the matrix, together with one internal tensorial parameter
being the matrix itself.

When a symbol for a mixing matrix is provided, it is mandatory to specify, in
addition, the name of a Les Houches block which will contain the numerical
values associated with the elements of the matrix. We indeed recall that both
\feynrules\ and most of the interfaced Monte Carlo event generators order
the model parameters according to a structure inspired by the Supersymmetry
Les Houches Accord (SLHA) \cite{Skands:2003cj,Allanach:2008qq}. In our example,
we impose the real part of the elements of $U_w$ to be stored in
a Les Houches block \texttt{WEAKMIX} and their imaginary part in an automatically
created block \texttt{IMWEAKMIX}, \ie, a block of the same name 
with the prefix {\tt IM} appended.

Implementing model Lagrangians might require to
explicitly use one or several of the mixing matrices for some of 
the model interactions, as for the Minimal
Supersymmetric Standard Model where the CKM matrix is employed in the
superpotential \cite{Duhr:2011se}. In this case, the  
matrices must be declared according to the standard syntax presented in the
\feynrules\ manual, numerical values being provided as inputs. This
subsequently 
renders the attribute \texttt{Block\-Na\-me} of the mixing class obsolete and 
ignored by \feynrules. 
Contrary, mixing matrices automatically declared through a mixing declaration 
cannot be employed in La\-gran\-gi\-ans.

In the Standard Model, the CKM matrix $V_{\rm CKM}$ relates the left-handed down
quark gauge-eigen\-sta\-tes $d_L^0$ to the mass-eigenstates $d_L$ as
\be
  d_L^0 = V_{\rm CKM} \cdot d_L \ .
\ee
To be compliant with the syntax presented so far, a symbol for the
hermitian-conjugate matrix has to be created. To avoid such a complication, the
optional attribute \texttt{Inverse} can be used and set to \texttt{True}, which
enforces a relation among the mass and gauge bases given by
\begin{verbatim}
  GaugeBasis = MixingMatrix . MassBasis
\end{verbatim}

\subsection{More advanced cases}\label{sec:adv}
\subsubsection{Scalar/pseudoscalar splittings}

When neutral scalar fields are mixing, the gauge eigenstates in general split
into their real degrees of freedom so that one scalar
and one pseudoscalar mass basis are required. Consequently, a list of two bases 
is provided as argument of the \texttt{MassBasis} attribute, instead of a single
basis as in Section \ref{sec:mixdecl}. Consistently, the arguments of the
attributes \texttt{Value}, \texttt{BlockName}, \texttt{Mi\-xing\-Ma\-trix} and
\texttt{Inverse} are also upgraded to lists. The first
element of those lists always refers to the scalar fields, while the second one
is related to the pseudoscalar fields. It may appear that some of the elements
of those lists are irrelevant, as for instance when the scalar mixing matrix is
unknown (\texttt{MixingMatrix} and \texttt{BlockName} are used) and the
pseudoscalar mixing matrix is known (\texttt{Value} is used). The irrelevant
list components are in this case replaced by underscores, as illustrated with  
\begin{verbatim}
  Mix["scalar"] == { 
    MassBasis    -> { {h1, h2}, {a1, a2} }, 
    GaugeBasis   -> { phi1, phi2 }, 
    BlockName    -> { SMIX, _ }, 
    MixingMatrix -> { US, _ },
    Value        -> { _, ... }
  } 
\end{verbatim}
where the (pseudo)scalar mass-eigenstates are represented by the symbols
\texttt{h1} and \texttt{h2} (\texttt{a1} and \texttt{a2}). In this example, the
mixing matrix related to the scalar sector is denoted by \texttt{US} and is 
associated with the Les Houches block \texttt{SMIX}. Concerning the
pseudoscalar sector, a numerical mixing matrix is instead provided (in the
ellipses). A concrete example is given for the 
Two-Higgs-Doublet Model in Section \ref{sec:2HDM}. 

\subsubsection{Dirac and Weyl fermion mixings}\label{sec:fermionmix}
Several options are left to the user concerning the implementation of Dirac
fermions mixings. Either one single gauge basis is employed, so that \feynrules\
internally takes care of the chirality projectors that appear in the related mass
terms, or different particle classes can be used for the left-handed and
right-handed components of the fermions. In this case, the \texttt{GaugeBasis}
attribute refers to a list of two gauge bases instead of to a single basis. 
For both options, the arguments of the
attributes \texttt{Value}, \texttt{BlockName}, \texttt{Mi\-xing\-Ma\-trix} and
\texttt{Inverse} consist of lists, the first component being related to the
mixing of the left-handed fermions and the second one to the mixing of the
right-handed fermions. As for neutral scalar mixing, underscores are used for
irrelevant list elements.

Lagrangian mass terms for charged Weyl fermions are generically
written as
\be
   \Big(\psi_1^-, \ldots, \psi_n^-\Big)\ M\ \bpm \chi^+_1\\ \vdots \\ \chi^+_n
    \epm \ ,
\ee 
where $M$ stands for the mass matrix and $\psi_i$ and $\chi_i$ are Weyl
fermions which have been assigned an electric charge of $\pm 1$ for the sake of
the example. The diagonalization of the matrix $M$ proceeds through two unitary
rotations $U$ and $V$,
\be
   \bpm \tilde\psi^-_1\\ \vdots \\ \tilde\psi^-_n \epm  = U\ 
    \bpm \psi^-_1\\ \vdots \\ \psi^-_n\epm \quad\text{and}\quad
   \bpm \tilde\chi^+_1\\ \vdots \\ \tilde\chi^+_n \epm  = V\ 
    \bpm \chi^+_1\\ \vdots \\ \chi^+_n\epm\ , 
\ee
which introduces two mass bases. Therefore, all the attributes
\texttt{MassBasis}, \texttt{GaugeBasis}, \texttt{Value},
\texttt{Mi\-xing\-Ma\-trix}, and \texttt{Block\-Na\-me} now take lists as
arguments (with underscores included where relevant). The only extra rule to
obey to is that the first components of these lists are associated with one
of the two rotations and the second components with the second of them. An
example is provided in Section \ref{sec:MSSM}.

\subsection{Vacuum expectation value declarations}\label{sec:vev}
In realistic new physics models, the ground state of the theory is non-trivial
and fields must be shifted by their vacuum expectation value. Since
Lorentz invariance and electric charge conservation impose that only
electrically 
neutral scalar fields can get non-vanishing vacuum expectation values, 
only shifts of (electrically) 
neutral scalar fields are allowed to be included in the
mixing relations. This information is encompassed within the variable
\texttt{M\$vevs} which consists of a list of two-component elements. The first
one refers to an unphysical field while the second one is
the associated vacuum expectation value. For instance, the declaration of a
configuration where two fields \texttt{phi1} and \texttt{phi2} get non-vanishing
vacuum expectation values \texttt{vev1} and \texttt{vev2} could be performed as 
\begin{verbatim}
  M$vevs = { { phi1, vev1 }, { phi2, vev2 } }
\end{verbatim}
The vacuum expectation values \texttt{vev1} and \texttt{vev2} must be declared
as any other model parameter, as described in the \feynrules\ manual
\cite{Christensen:2008py}.

\subsection{User functions}\label{sec:usrfct}
Once both the mixing relations and the vacuum expectation values 
have been properly declared, the mass matrices of the model can be extracted by
means of the function \texttt{ComputeMassMatrix}, 
\begin{verbatim}
 ComputeMassMatrix[ Lag, options ] 
\end{verbatim}
where \texttt{Lag} is the model Lagrangian and the symbol \texttt{options}
stands for optional arguments. If no option is provided, the function calculates
all the mass matrices of the model for which the numerical value of the mixing
matrix is unknown. It is possible to focus on a specific mixing relation whose
label is denoted by \texttt{"l1"} by issuing, in \mathematica, 
\begin{verbatim}
 ComputeMassMatrix[ Lag, Mix->"l1" ] 
\end{verbatim}
For the computation of multiple matrices, the label 
{\tt "l1"} has to be replaced by a list of labels.
During the computation of the mass matrices, a
lot of information is by default printed to the screen. This can be avoided by
including the optional argument \texttt{Screen\-Out\-put -> False} in the two
command lines above.

The input information and the result of the
\texttt{Com\-pu\-te\-Mass\-Ma\-trix} function can be retrieved
through the intuitive printing
functions, \texttt{MassMatrix}, \texttt{Gau\-ge\-Ba\-sis}, \texttt{MassBasis},
\texttt{MixMatrix}, \texttt{BlockName} and \texttt{Ma\-trix\-Sym\-bol} which all
take as argument the label of a mixing relation. A wrapper is also available,
\begin{verbatim}
  MixingSummary [ "l1" ]
\end{verbatim}
which sequentially calls all the printing functions for a mixing relation
represented by the label \texttt{"l1"} and organizes the output in a readable
form.

The \feynrules\ method to extract analytically a mass matrix is fully
generic and can 
be employed to compute any matrix $M$ defined by the Lagrangian  
\be
  {\cal L}_{\rm mass} = {\cal B}_2^\dag \ M\  {\cal B}_1 \ ,
\ee
where ${\cal B}_1$ and ${\cal B}_2$ stand for two field bases possibly different. 
The calculation
of the matrix $M$ is achieved by issuing 
\begin{verbatim}
  ComputeMassMatrix[ Lag, 
      Basis1 -> b1, Basis2 -> b2 ]
\end{verbatim}
where the symbols \texttt{b1} and \texttt{b2} are associated with the bases
${\cal B}_1$ and ${\cal B}_2$ and refer to lists of fields. 
In this case, the printing functions introduced above are not
available.


\section{Automated spectrum generation}
\label{sec:ass}

\subsection{The \nc\ package}\label{sec:asperge}

The computation of the unknown mixing matrices necessary for diagonalizing all the model
mass matrices can in general only be achieved numerically. To this end, we have
developed the \cpp\ program \nc.
It includes a set of \cpp\ source files (stored in the
subdirectory \texttt{src}), coming together with the related header files
(sto\-red in the subdirectory \texttt{inc}), that can be split into
model-independent and model-dependent files. For an efficient 
use of the \nc\ program, it has been entirely embedded within the \feynrules\
package. Therefore, only a brief discussion of the structure of the code is
presented in this paper. More information, such as a {\sc doxygen}
documentation, can be found on the \nc\ webpage \cite{web:asperge}.

The set of model-independent files contains, on the one hand, several tools
dedicated to matrices and their diagonalization
(\texttt{MassMatrix.cpp}, \texttt{Mass\-Ma\-trix.hpp} as well as
\texttt{Matrix.hpp}). 
On the other hand, the \nc\ code is based on an internal format 
for parameters, defined in the source files \texttt{Par.cpp}, \texttt{CPar.cpp},
\texttt{RPar.cpp} and in the associated header files. This format is
inspired from a SLHA structure and the corresponding mapping is encoded into
the files \texttt{ParSLHA.cpp}, \texttt{SLHA\-Block.cpp}, and in the
associated header files. Finally, printing and string manipulation routines are
included in the files \texttt{tools.cpp} and \texttt{tools.hpp} and the
program comes with a makefile.

All the model dependency is included in the two files \texttt{Parameters.cpp}
and \texttt{Parameters.hpp} as well as in the core program implemented in the
\texttt{main.cpp} file. 

The information encompassed in the two parameter files
is threefold. First, the SLHA structure ordering the external parameters is
encoded in terms of blocks and counters. Next, the definitions of the internal
parameters as functions of the other model parameters are implemented, where a
proper running of the \nc\ program is only guaranteed if the parameters do not
depend on the masses and mixing matrices to be computed. Finally, the analytical
formulas of the mass matrices to diagonalize are included. 

The main program (\texttt{main.cpp}) starts with the declaration of the
different mass matrices of the model. Links to the relevant elements of the 
mass basis are then implemented by means of the associated PDG codes, which 
allow to assign the mass eigenvalues to each of the physical particles, the
ordering of the PDG codes following the mass ordering.

\subsection{Interfacing the \nc\ package to \feynrules}
\label{sec:interface}

The \nc\ package can be entirely generated, for a given
particle physics model, from the \feynrules\ model information by means of a
dedicated interface which works as for the other \feynrules\
interfaces. It is then called by typing, in a \mathematica\ session,
\begin{verbatim}
  WriteASperGe[ Lag, Output -> dirname ] 
\end{verbatim}
where the symbol \texttt{Lag} stands for the model Lagrangian and
\texttt{Output->dirname} for an optional argument indicating the name of the
directory where to store all the created files. If
unspecified, the directory \texttt{ModelName\textunderscore MD} is employed,
\texttt{Mo\-del\-Na\-me} being the name of the \feynrules\ model. 

The interface first extracts all the relevant mass matrices from the Lagrangian
\texttt{Lag} by means of the function
\texttt{Com\-pu\-te\-Mass\-Ma\-trix} introduced in Section
\ref{sec:usrfct}. It then writes, in addition to model-independent files
described in Section \ref{sec:asperge}, the three model-dependent files 
\texttt{main.cpp}, \texttt{Parameters.cpp} and \texttt{Parameters.hpp}, together
with one data file \texttt{Ex\-ter\-nals.dat} (stored in the
subdirectory \texttt{input}). This last file contains the numerical values of
the external 
parameters of the model, necessary for the numerical evaluation of the mass
matrices. When running the code (see Section \ref{sec:running}), the user can
update this file or even employ a different file according to his needs.

The numerical matrix diagonalization performed by \nc\ is based on \gsl\
functions relying
on the hermiticity of the mass matrices which employs
symmetric bi-diagonalization followed by QR reduction.
This contrasts with existing diagonalization packages developed in the framework of 
\feynarts\ \cite{Hahn:2006hr} and \calchep\ \cite{Belanger:2010st} that are 
based on Jacobi-type iterative algorithms. A hermiticity check is therefore performed by the
interface before writing down the output. Since the mass matrix $M$ related to 
charged fermions is by construction non-her\-mi\-ti\-an, the matrices $M^{\dagger} M$
and $M M^{\dagger}$ are employed instead, which allows to obtain left-handed and
right-handed fermion mixing matrices separately. 

It is also possible to focus on one or several 
specific mixing relations. In this case, the {\tt Mix} option, already introduced
in the context of the {\tt ComputeMassMatrix} function, has to be used, 
\begin{verbatim} 
  WriteASperGe[ Lag, Mix -> {"l1", "l2"} ]
\end{verbatim} 
We refer 
to Section \ref{sec:usrfct} for more information.

\subsection{Running \nc}\label{sec:running}
Since the \nc\ package is based on \gsl\ functions, 
it is mandatory to have the \gsl\ libraries installed on the system. Then,
if the \texttt{g++} compiler is available, the makefile generated by
\feynrules\ can be employed directly. Otherwise, it must 
be first edited accordingly to include proper compiler information.

Once compiled, \nc\ can be executed by typing in a shell
\begin{verbatim}
  ./ASperGe <infile> <outfile>
\end{verbatim}
where the arguments indicate in which file  the numerical value of the external
parameters must be read (\texttt{<in\-file>}) and where to store the output file
(\texttt{<outfile>}). This file contains, in addition to the input parameters,
the computed numerical values of the 
mixing matrices, split in terms of their real and imaginary parts according to
the SLHA conventions, as well as all the masses of the physical states (stored
in the SLHA block \texttt{MASS}). In order to execute \nc\ with
all the default settings as generated by the \feynrules\ interface, it is
sufficient to type in a shell
\begin{verbatim}
 ./ASperGe input/externals.dat output/out.dat
\end{verbatim}

Both the compilation and the execution of the program can be performed from the 
\mathematica\ session, by issuing
\begin{verbatim}
  RunASperGe[ ]
\end{verbatim}
This also loads the SLHA parameter file \texttt{out.dat} back into the
\feynrules\ session, so that it can be further employed, \eg, to generate a UFO
model. Information about the run of \nc\ can be found
in the file \texttt{ASperGe.log} stored in the same folder as the
executable. 

It is also possible to diagonalize specific mass
matrices of the model by executing 
\begin{verbatim}
  ./ASperGe <infile> <outfile> m1 m2 ...
\end{verbatim}
where \texttt{m1}, \texttt{m2}, \etc, are the names of the mixing matrices to be computed.


\section{Illustrative examples}\label{sec:example} 
In this section, we illustrate the features of the \nc\ program and its
interface to \feynrules\ by choosing three extensions of the Standard Model 
with non-trivial mixing relations, \ie, the Two-Higgs-Dou\-blet Mo\-del (2HDM),
the Minimal Left-Right Symmetric Standard Mo\-del (LRSM) and the Minimal Supersymmetric
Standard Model (MSSM). We modify their original \feynrules\ 
implementations\footnote{Since no previous 
implementation of the LRSM exists, we take the opportunity
to provide the relevant details in Section~\ref{sec:LRSM}.}
\cite{Christensen:2009jx,Duhr:2011se}
to accommodate for the mixings as described in Section \ref{sec:technicalFR}.
We then employ the \nc\ program (see Section \ref{sec:ass}) to numerically 
calculate some of the mass and mixing matrices of these models.

\subsection{The general Two-Higgs-Doublet Model}\label{sec:2HDM}
The 2HDM is one of the simplest extensions of the Standard Model, with
respect to which it only contains 
a second weak doublet of scalar fields. Following
the conventions of the original \feynrules\ implementation \cite{Christensen:2009jx},
both Higgs fields $\phi_1$ and $\phi_2$ carry the same hypercharge so that they 
can always be redefined by means of $U(2)$ transformations 
\cite{Branco:1999fs,Ginzburg:2004vp,Davidson:2005cw,Haber:2006ue}. 
Adopting the so-called Higgs-basis, the two doublets read
\renewcommand{\arraystretch}{1.2}%
\be
\phi_1 =
  \begin{pmatrix}
    G^+ \\ \frac{ v + H^0 + i G^0}{\sqrt{2}} 
  \end{pmatrix}  
 \qquad\text{and} \qquad 
\phi_2 =
  \begin{pmatrix}
    H^+ \\ \frac{R^0 + i I^0}{\sqrt{2}} 
  \end{pmatrix} \ ,
\label{eq:hdoublet} \ee \renewcommand{\arraystretch}{1.}%
where only the neutral component of the $\phi_1$ field acquires a vacuum 
expectation value $v$. Moreover, the 
Goldstone bosons $G^\pm$ and $G^0$ as well as the charged Higgs field $H^\pm$ are 
not required to be further rotated, so that only  
the mass matrix of the neutral fields $H^0$, $R^0$ and $I^0$ must still 
be diagonalized.

We extend the 2HDM \feynrules\ implementation described in Ref.\ 
\cite{Christensen:2009jx} by first indicating that the second component of the 
$\phi_1$ field, represented by the symbol {\tt phi1}, acquires a 
non-vanishing vacuum expectation value $v$ labeled by the symbol {\tt vev}, 
\begin{verbatim}
  M$vevs = { {phi1[2], vev} }
\end{verbatim}
as shown in Section \ref{sec:vev}.
Then, we choose to implement the mixing of the $H^0$, $R^0$ and $I^0$ fields to
the physical $h_1$, $h_2$ and $h_3$ fields in a two-step manner. In a first stage, 
the gauge eigenstates are split into their scalar and pseudoscalar components, 
\begin{verbatim}
  Mix["1p"] == {
    MassBasis  -> { {H0}, {G0} },
    GaugeBasis -> { phi1[2] },
    Value      -> { {{1}}, {{1}} }
  }

  Mix["2p"] == {
    MassBasis  -> { {R0}, {I0} },
    GaugeBasis -> { phi2[2] },
    Value      -> { {{1}}, {{1}} }
  }
\end{verbatim}
following the syntax introduced in Section \ref{sec:mixdecl} and Section \ref{sec:adv}
and making use of 
the self-explained symbols {\tt H0} and {\tt R0} ({\tt G0} and {\tt I0})
for representing the (pseudo)scalar degrees of freedom.
Similarly, we can employ the mixing infrastructure to map the 
charged components of $\phi_1$ and $\phi_2$ to the physical fields 
$G^+$ and $H^+$ by means of a $1\times 1$ identity matrix. Since this procedure is trivial, we
omit any further details from the present manuscript and refer to the model 
implementation~\cite{web:asperge}. In a second stage, 
the rotation to the physical fields, represented by the symbols {\tt h1}, {\tt h2} 
and {\tt h3}, is declared as 
\begin{verbatim}
  Mix["1s"] == {
    MassBasis    -> { h1, h2, h3 },
    GaugeBasis   -> { H0, R0, I0 },
    MixingMatrix -> NH,
    BlockName    -> NHMIX
  }
\end{verbatim}
where we associate the symbol {\tt NH} to the corresponding mixing matrix and assign
the Les Houches block ({\tt IM}){\tt NHMIX} to the numerical value of its elements.
The neutral squared mass matrix ${\cal M}^2$ can then be derived from the model 
Lagrangian (represented by the symbol {\tt L2HDM}) by typing, in the \mathematica\ session,
\begin{verbatim}
 ComputeMassMatrix[L2HDM, Mix->"1s"]
\end{verbatim}
As a result, one recovers the well-known expression depending on the most general
scalar potential parameters $\lambda_i$ and $\mu_i$ (see Ref.~\cite{Christensen:2009jx} 
for further information), \renewcommand{\arraystretch}{1.2}%
\be\bsp
  & {\cal M}^2 = \\
  &\! \begin{pmatrix}
    2 \lambda_1 v^2 & \Re[\lambda_6] v^2 & - \Im[\lambda_6]v^2 \\
    \Re[\lambda_6] v^2 & m_{\pm}^2 \!+\! \Big[\frac{\lambda_4}{2} \!+\! \lambda_5\Big] v^2 & 0 \\
   - \Im[\lambda_6]v^2 & 0 & m_{\pm}^2 \!+\! \Big[\frac{\lambda_4}{2} \!-\! \lambda_5\Big]v^2
  \end{pmatrix}\ , 
\esp\ee \renewcommand{\arraystretch}{1.0}%
after having introduced the squared mass of the charged Higgs boson 
$m_{\pm}^2 = 1/2 \lambda_3 v ^2 + \mu_2$ and 
removed two of the $\mu$-parameters by means of the potential minimization conditions, 
$\mu_1 = -\lambda_1 v^2$ and $\mu_3 = - 1/2 \lambda_6 v^2$. 

The numerical value of the unitary matrix $U$ diagonalizing ${\cal M}^2$ is 
obtained by generating and making use of the \nc\ package, as shown in 
Section \ref{sec:ass}. 
We fix, adopting a representative benchmark scenario, the Higgs potential parameters to 
$\lambda_1 = \lambda_2 = \lambda_3 = 1.0$, $\lambda_4 = 0.5$, $\lambda_5 = 0.4$, 
$\lambda_6 = 0.3$, $\lambda_7 = 0.2$ and $\mu_2 = 6 \cdot 10^{4}$~GeV. This leads to the 
three mass eigenvalues
\be
  ( m_{h1}, m_{h2}, m_{h3} ) = (285, 327, 379)  \text{ GeV,}
\ee 
whereas the mixing matrix reads	
\be
  U = \begin{pmatrix}
    0     &\  0    &\ \ -i \\
    \phantom{-}0.784  &\ -0.621 &\ \ \phantom{-}0  \\
    -0.621  &\ -0.784 &\ \ \phantom{-}0
  \end{pmatrix} \ .
\ee
These results are in good agreement with those obtained by means of the 
{\sc TwoHiggsCalc}\ calculator \cite{Alwall:2007st}.

\subsection{The Minimal Left-Right Symmetric Standard Model}\label{sec:LRSM}

\begin{table}[!t]
\centering
\caption{Field content of the LRSM, given together with their representation under
the $SU(3)_c \times SU(2)_L \times SU(2)_R  \times U(1)_{B-L}$ gauge group. The 
$SU(2)_L$ ($i,j=1,2$) and $SU(2)_R$ ($i',j'=1,2$) fundamental 
index structure is explicitly indicated.}
\label{tab:lrsmcontent} \renewcommand{\arraystretch}{1.2}%
\begin{tabular*}{\columnwidth}{@{\extracolsep{\fill}}lll@{}}
\hline
 Field& Components & Representation \\ 
\hline
   $Q_L^i$ & 
   $\begin{pmatrix} u_L\\ d_L \end{pmatrix}$  & 
   $(\utilde {\bf 3}, \utilde {\bf 2}, \utilde {\bf 1}, \frac13)$ \\
   $Q_{R i'}$ & 
   $\begin{pmatrix} u_R^c & d_R^c \end{pmatrix}$  & 
   $(\utilde {\bf \bar 3},  \utilde {\bf 1}, \utilde {\bf 2}^*, -\frac13)$ \\
   $L_L^i$ & 
   $\begin{pmatrix} \nu_L\\ \ell_L \end{pmatrix}$  & 
   $(\utilde {\bf 1}, \utilde {\bf 2}, \utilde {\bf 1}, -1)$ \\
   $L_{R i'}$ & 
   $\begin{pmatrix} \nu_R^c & \ell_R^c \end{pmatrix}$  & 
   $(\utilde {\bf 1},  \utilde {\bf 1}, \utilde {\bf 2}^*, 1)$ \\
\hline
   $\Phi^i{}_{i'}$ &
   $\begin{pmatrix} \Phi^0 & \Phi^+ \\ \Phi^{'\prime -} & \Phi'{}^0 \end{pmatrix}$ & 
   $(\utilde {\bf 1} , \utilde {\bf 2}, \utilde{\bf 2}^*, 0)$\\
   $\Delta_L{}^i{}_j$ & 
   $\begin{pmatrix} \frac1{\sqrt{2}} \Delta_L^+ & \Delta_L^{++} \\
     \Delta_L^0& -\frac1{\sqrt{2}} \Delta_L^+\end{pmatrix}$ &
   $(\utilde {\bf 1} , \utilde {\bf 3}, \utilde {\bf 1}, 2)$ \\
   $\Delta_R{}^{i'}{}_{j'}$& 
   $\begin{pmatrix} \frac1{\sqrt{2}} \Delta_R^+ & \Delta_R^{++} \\
      \Delta_R^0& -\frac1{\sqrt{2}} \Delta_R^+ \end{pmatrix}$ & 
   $(\utilde {\bf 1} , \utilde {\bf 1}, \utilde {\bf 3}, 2)$\\
\hline
\end{tabular*}\renewcommand{\arraystretch}{1.0}%
\end{table}

The LRSM \cite{Pati:1974yy,Mohapatra:1974hk,Mohapatra:1974gc,Senjanovic:1975rk,%
Mohapatra:1977mj,Senjanovic:1978ev,Lim:1981kv} is an extension of the Standard Model with an
enlarged $SU(3)_c \times SU(2)_L \times SU(2)_R 
\times U(1)_{B-L}$ gauge symmetry. In this model, the fermionic degrees of 
freedom of the Standard Model lying in the trivial representation of $SU(2)_L$ are
collected into $SU(2)_R$ doublets, as shown in the first part of Table \ref{tab:lrsmcontent}
where the model matter field content is presented together with 
the associated quantum numbers.
In addition, the symmetry-breaking mechanism down to electromagnetism is also
more involved, relying on an enriched Higgs sector (see the second part of the table).

The LRSM Lagrangian consists of standard kinetic and gauge interaction terms
for all fields as well as of the Yukawa interactions 
\be\bsp
{\cal L}_{\text{LR}}^{\text{Y}} =&\
   \bar{Q}^c_L {\bf y_Q^{(1)}} \hat\Phi Q_R +
   \bar{L}^c_L {\bf y_\ell^{(1)}} \hat \Phi L_R 
\\ &\
   + \bar{Q}^c_R {\bf y_Q^{(2)}}  \Phi^\dag Q_L +
   \bar{L}^c_R {\bf y_\ell^{(2)}} \Phi^\dag L_L \\
&\ + \hat{\bar{L}}^c_L {\bf y_\ell^{(3)}} \Delta_L L_L + 
   \hat{\bar{L}}_R {\bf y_\ell^{(4)}} \Delta_R L_R^c + \text{h.c.} \ .
\esp \ee
In this equation, all indices are understood, 
the matrices ${\bf y_Q^{(i)}}$ and ${\bf y_\ell^{(i)}}$ 
are $3\times 3$ matrices in flavor space
and the superscript $c$ indicates charge conjugation\footnote{We recall that
the components of the field $Q_R$ are charge conjugate (see
Table \ref{tab:lrsmcontent}).}.
Moreover, gauge invariance is ensured by the introduction of the hatted fields
\be
 \hat \Phi_i{}^{i'} = \epsilon_{ij} \epsilon^{i'j'}\Phi^j{}_{j'} \ , \
 \hat L_{Li} = \epsilon_{ij} L_L^j \ , \
 \hat L_R^{i'} = \epsilon^{i'j'} L_{Rj'} \ ,
\ee
where the rank-two antisymmetric tensors with
lower and upper indices are defined by $\epsilon_{12}=-\epsilon^{12}=1$. 
Introducing Higgs mass parameters $\mu_i$ and quartic interaction strengths
$\lambda_i$, $\rho_i$ and $\alpha_i$, the scalar potential reads

\begin{widetext}
\be\bsp
  & {\cal L}^{\rm H} =
    \mu_1^2 \Tr[\Phi^\dag \Phi] 
  - \lambda_1 \Big(\Tr[\Phi^\dag \Phi]\Big)^2
  - \lambda_2 \Tr[\Phi^\dag \Phi\Phi^\dag \Phi] 
  - \frac12 \lambda_3 \Big( \Tr[\hat\Phi\Phi^t] + \Tr[\Phi^\dag \hat\Phi^{\dag t}]\Big)^2
  - \lambda_4 \Tr[\Phi^\dag \Phi \hat\Phi^t\hat\Phi^{\dag t}] 
\\&\ 
  - \frac12 \lambda_5 \Big( \Tr[\hat\Phi\Phi^t]  - \Tr[\Phi^\dag \hat\Phi^{\dag t}] \Big)^2
  - \frac12 \lambda_6 \Big( \Tr[\Phi^\dag \hat\Phi^{\dag t} \Phi^\dag \hat\Phi^{\dag t}] + 
         \Tr[\hat\Phi^t \Phi \hat\Phi^t \Phi]\Big)
  + \mu_2^2 \Big(\Tr[\Delta_L^\dag \Delta_L] + \Tr[\Delta_R^\dag\Delta_R] \Big) 
\\&\
  - \rho_1 \Big( \Tr[\Delta_L^\dag\Delta_L]^2 + \Tr[\Delta_R^\dag\Delta_R]^2\Big) 
  \!-\! \rho_2 \Big( \Tr[\Delta_L^\dag \Delta_L\Delta_L^\dag \Delta_L] \!+\! 
     \Tr[\Delta_R^\dag \Delta_R\Delta_R^\dag \Delta_R]\Big) 
  \!-\! \rho_3 \Tr[\Delta_L^\dag \Delta_L] \Tr[\Delta_R^\dag \Delta_R]
\\&\
  - \alpha_1 \Tr[\Phi^\dag \Phi] \Big(\Tr[\Delta_L^\dag \Delta_L] + \Tr[\Delta_R^\dag \Delta_R]\Big)
  - \alpha_2 \Big(\Tr[\Delta_R^\dag \Phi^\dag \Phi \Delta_R] + 
      \Tr[\Delta_L^\dag \Phi \Phi^\dag \Delta_L]\Big)
\\&\ 
 - \alpha_3 \Big(\Tr[\Delta_R^\dag \hat\Phi^t \hat\Phi^{\dag t}\Delta_R] + 
     \Tr[\Delta_L^\dag \hat\Phi^{\dag t}\hat\Phi^t \Delta_L]\Big) \ . 
\esp\ee
\end{widetext}
\noindent 
\noindent Since the corresponding \feynrules\ model
description is standard, we refer to the \feynrules\ manual \cite{Christensen:2008py} and 
leave all implementation details out of this work. 
 
In the LRSM, the symmetry-breaking mechanism is performed in two steps.
At high energy, the $SU(2)_L \times SU(2)_R
\times U(1)_{B-L}$ gauge symmetry is spontaneously broken to the electroweak symmetry,
the latter being subsequently broken to electromagnetism at 
a lower scale. Consequently, the neutral components of the sca\-lar fields get
vacuum expectation values at the minimum of the potential, 
$\langle \Phi^0\rangle = v/\sqrt{2}$, $\langle \Phi^{\prime 0}\rangle = 
v'/\sqrt{2}$ and $\langle\Delta_{L,R}^0\rangle = v_{L,R}/\sqrt{2}$, by which they are shifted.
In the rest of this section, we focus on the mixing of the neutral Higgs fields and
illustrate the way to implement a two-stage field rotation.
For all the other mixing relations of the LRSM, we refer to the 
implementation~\cite{web:asperge}.

We first assume, motivated by neutrino mass and kaon system data 
\cite{Lim:1981kv,Mohapatra:1980yp}, that  $v_L = v' \approx 0$. 
Next, we implement the rotation associated with the 
diagonalization of the third generator of $SU(2)$ in the adjoint representation as 
\begin{verbatim}
  Mix["2a"] == {
    MassBasis  -> { DLpp, DL0 }, 
    GaugeBasis -> { DL[1], DL[2] }, 
    Value      -> { {1/Sqrt[2], -I/Sqrt[2]},
                    {1/Sqrt[2],  I/Sqrt[2]}}
  }
\end{verbatim}
depicting the 
example of the $SU(2)_L$ Higgs triplet.
The\-se replacement rules translate the rotation of the $\Delta_L^1$ and $\Delta_L^2$ fields,
represented by the {\tt DL[1]} and {\tt DL[2]} symbols, to the 
$\Delta^0$ and $\Delta^{++}$ states labeled by {\tt DL0} and {\tt DLpp}. 
Then, the neutral fields 
$\Delta_L^0$, $\Delta_R^0$, $\Phi^0$ and $\Phi^{\prime 0}$, represented
by the symbols {\tt DL0}, {\tt DR0}, {\tt phi[1,1]} and {\tt phi[2,2]}, mix
to four scalar degrees of freedom $h^0_1$,  $h^0_2$,  $h^0_3$ and $h^0_4$, 
two physical pseudoscalar Higgs bosons $a^0_1$ and $a^0_2$ and 
two Goldstone bosons $G^0_1$ and $G^0_2$ to be eaten by the $Z$ and $Z'$ 
vector fields when getting massive.
Introducing the corresponding symbols {\tt h01}, {\tt h02}, {\tt h03}, {\tt h04}, 
{\tt a01}, {\tt a02}, {\tt G01} and {\tt G02}, these rotations are implemented as
\begin{verbatim}
  Mix["2e"] == {
    MassBasis    -> { {h01,h02,h03,h04},
                      {G01,G02,a01,a02} }, 
    GaugeBasis   -> 
      { DL0,DR0,phi[1,1],phi[2,2] }, 
    MixingMatrix -> { UHN,UAN }, 
    BlockName    -> { HMIX,AMIX} 
  }
\end{verbatim}
The two symbols {\tt UHN}
and {\tt AUN} respectively denote the scalar and pseudoscalar mixing matrices, the 
numerical value
of their elements being included in the two Les Houches blocks ({\tt IM}){\tt HMIX} and 
({\tt IM}){\tt AMIX}.

Typing, in \mathematica, the commands 
\begin{verbatim}
 ComputeMassMatrix[Lag, Mix->"2e"]
 MassMatrix["2e", "S"]
\end{verbatim}
allows to calculate both the scalar and pseudoscalar mass matrices and display
the scalar squared mass matrix $\tilde{\cal M}^2$ to the screen. It reads,
\be
 \tilde{\cal M}^2 = \bpm 
  A & 0 & 0 & 0 \\
  0 & B & \big(\alpha_1+\alpha_3\big) v\ v_R & 0 \\
  0 & \big(\alpha_1+\alpha_3\big) v\ v_R & C & 0 \\
  0 & 0 & 0 & D 
 \epm \ ,
\ee 
where we have introduced the quantities
\be\bsp
  A =&\ \frac12 (\alpha_1 + \alpha_3) v^2 - \mu_2^2 + \frac12 \rho _3 v_R^2 \ ,\\ 
  B =&\ \frac12 (\alpha_1 + \alpha_3) v^2 - \mu_2^2  +3 (\rho _1 - \rho _2) v_R^2\ , \\ 
  C =&\ 3 (\lambda_1 + \lambda_2) v^2 + \frac12 (\alpha_1 +\alpha_3) v_R^2 - \mu_1^2\ , \\
  D =&\ (\lambda_1 \!+\! 4 \lambda_3 \!+\! \lambda_4 \!+\! \lambda_6) v^2 +
        \frac12(\alpha_1 \!+\! \alpha_2) v_R^2 - \mu_1^2 \ . 
\esp\ee

In order to numerically compute the unitary matrix $U$ diagonalizing $\tilde{\cal M}^2$, we use
the \nc\ program, generated from \feynrules\ by issuing 
\begin{verbatim}
    WriteAsperge[Lag]
\end{verbatim}
after fixing the external Lagrangian parameters of the model to $\lambda_1 = \lambda_2 =  
\lambda_3 = \lambda_4 = \lambda_6 = 0.1$, $\alpha_1 = 0.1$, $\alpha_2 = 0.3$,
$\alpha_3 = 0.1$, $\rho_1 = 0.1$, $\rho_2 = 0$ and $\rho_3 = 0.5$. In addition,
the $\mu$-terms are deduced from the minimization conditions of the scalar potential and
the vacuum expectation values are taken as 
$v = 248$~GeV and $v_R = 6000$~GeV. It should be noted that the relevance of such numerical
values is going beyond the scope of this paper, and could be addressed by means of external
packages such as the one presented in Ref.~\cite{Coimbra:2013qq}.
Once \nc\ is executed, one obtains 
\be
  U = \bpm
    0 & -0.041 & 0.99 & 0 \\
    0 & 0 & 0 & 1 \\
    1 & 0 & 0 & 0 \\
    0 & 0.99 & 0.041 & 0 
  \epm \ ,
\ee
the mass eigenvalues being
\be
  (m_1, m_2, m_3, m_4) = (111, 1905, 2324, 2686)~\text{GeV.}
\ee

\subsection{The Minimal Supersymmetric Standard Model}\label{sec:MSSM}
In supersymmetric extensions of the Standard Model, each of the model's degrees of freedom 
comes accompanied by  
a superpartner with opposite statistics. The minimal version of such theories, the so-called
MSSM \cite{Nilles:1983ge,Haber:1984rc}, 
has been originally implemented in \feynrules\ by making use of its superspace module
\cite{Duhr:2011se}. We hence refer, on the one hand, to Ref.\ \cite{Duhr:2011se} for notations
and conventions and, on the other hand, to the new model implementation \cite{web:asperge} for more 
information on the way in which particle mixings have been implemented.  
In the rest of this subsection, we employ the MSSM implementation to illustrate 
fermion mixing declaration.

After electroweak symmetry breaking, the left-hand\-ed, two-component Weyl fermionic, 
gaugino and higgsino 
fields $\tilde W^\pm$ and $\tilde H_d^- / \tilde H_u^+$ 
mix to the chargino eigenstates $\chi^\pm$. Introducing
the two rotation matrices $U$ and $V$ (labeled by the symbols {\tt UU} and {\tt VV}), 
this mixing is declared through an instance of the mixing class, 
\begin{verbatim}
  Mix["3d"]=={
    MassBasis    -> { {chmw[1],chmw[2]}, 
                      {chpw[1],chpw[2]} }, 
    GaugeBasis   -> { {wowm,hdw[2]},
                      {wowp,huw[1]} }, 
    BlockName    -> {UMIX,VMIX}, 
    MixingMatrix -> {UU,VV} 
  }
\end{verbatim}
In this list of replacement rules, {\tt wowm} ({\tt hdw}) and {\tt wowp} ({\tt huw}) are 
the labels of the negatively and positively 
charged wino (higgsino) states, the related mass eigenstates being represented
by the symbols {\tt chmw} and {\tt chpw}. In addition, we link the mixing matrices $U$ and $V$ 
to the Les Houches blocks {\tt UMIX} and {\tt VMIX}.  

Like in the previous subsections, we employ \feynrules\ to 
extract the corresponding tree-level mass matrix ${\cal M}'$ from the Lagrangian,
\be
  {\cal M}' = \bpm
    M_2&\sqrt2 m_W \sin \beta\\
    \sqrt2 m_W \cos \beta& \mu
  \epm \ ,
\ee
where $m_W$ denotes the $W$-boson mass, $\mu$ the superpotential Higgs mixing parameter, $M_2$
the supersymmetry-breaking wino mass and $\tan\beta$ is defined as the ratio of the two neutral
Higgs field vacuum expectation values $\tan\beta = \langle H_u^0/H_d^0\rangle$.

As in the original implementation, we choose the typical minimal supergravity point SPS 1a 
\cite{Allanach:2002nj} as a benchmark scenario. Then, we generate the MSSM \nc\ program and
use it to compute numerically the mixing matrices $U$ and $V$, 
\be 
  U\! =\! \bpm
    0.918 & -0.397 \\
    -0.397 & -0.918
\epm \ ,\ \ 
  V \!=\! \bpm
    0.974 & -0.226 \\
    -0.226 & -0.974
\epm \ , 
\ee
as well as the corresponding mass eigenvalues 
\be
  (m_{\chi_1^\pm}, m_{\chi_2^\pm}) = (176, 382) \text{ GeV.}
\ee
Those values are in good agreement with those returned by commonly used 
MSSM spectrum generators.


\section{Summary}\label{sec:summary}
In this paper, we have presented an extension of the \feynrules\ package dedicated
to the automated generation of the particle mass spectrum and mixing
structure associated to any Lagrangian-based quantum field theory. The new
module is based on the introduction of a new structure for particle mixing
declaration allowing, on the one hand, for the analytical computation of all the model 
mass matrices, and, on the other hand, for the generation of a \cpp\ program
dubbed \nc, yielding the numerical evaluation of the associated rotation matrices. We illustrate
the strength of this new module in the context of the Two-Higgs-Doublet Model, the Minimal
Left-Right Symmetric Standard Model and the Minimal Supersymmetric Standard Mo\-del.


\begin{acknowledgements}
The authors are grateful to N.D.\ Christensen, C.\ Degrande and C.\ Duhr for useful discussions
on the project. This work has been partially supported by a Ph.D.\ fellowship of the French 
ministry for education and research, by the Theory-LHC
France-initiative of the CNRS/IN2P3, by the French ANR 12 JS05 002 01 BATS@LHC,  
by the Concerted Research action `Supersymmetric
Models and their Signatures at the Large Hadron Collider' and the Strategic Research Program
`High Energy Physics' of the Vrije Universiteit Brussel (VUB), by the Belgian Federal Science Policy Office through the Interuniversity Attraction Pole IAP VI/11 and P7/37 and by a `FWO-Vlaanderen' aspirant fellowship.

\end{acknowledgements}

\end{document}